# SAPHIR - a multi-scale, multi-resolution modeling environment targeting blood pressure regulation and fluid homeostasis

S. Randall Thomas, Enas Abdulhay, Pierre Baconnier, Julie Fontecave, Jean-Pierre Françoise, François Guillaud, Patrick Hannaert, Alfredo Hernandez, Virginie Le Rolle, Pierre Maziere, Fariza Tahi, Farida Zehraoui

*Abstract*—We present progress on a comprehensive, modular, interactive modeling environment centered on overall regulation of blood pressure and body fluid homeostasis. We call the project SAPHIR, for "a Systems Approach for PHysiological Integration of Renal, cardiac, and respiratory functions". The project uses state-of-the-art multi-scale simulation methods. The basic core model will give succinct input-output (reduced-dimension) descriptions of all relevant organ systems and regulatory processes, and it will be modular, multi-resolution, and extensible, in the sense that detailed sub-modules of any process(es) can be "plugged-in" to the basic model in order to explore, eg. system-level implications of local perturbations. The goal is to keep the basic core model compact enough to insure fast execution time (in view of eventual use in the clinic) and yet to allow elaborate detailed modules of target tissues or organs in order to focus on the problem area while maintaining the system-level regulatory compensations.

## I. INTRODUCTION

DESPITE the development of ever more sophisticated models at many scales, from the level of gene regulation or intra-cellular signal transduction or cellular metabolism, through models of tissue function, epithelial transport, or even whole organ physiology, there is presently, no comprehensive, organism-level modeling environment (i.e., open for inspection, modification, & extension) that allows exploration of the whole chain of regulatory influences brought into play by a local perturbation such as a defect in a cell membrane ion transport protein.

In this paper, we present progress on a comprehensive, modular, interactive modeling environment centered on overall regulation of blood pressure and body fluid homeostasis. The project uses state-of-the-art multi-scale simulation methods to construct a basic, modular core model whose components can be replaced individually by higher resolution versions to address specific questions at the organ or tissue level without abandoning the systems-level feedback regulation. The basic core model gives succinct input-output (reduced-dimension) descriptions of all relevant organ systems and regulatory processes, and is modular, multi-resolution, and extensible, in the sense that detailed sub-modules of any process(es) can be "plugged-in" to the basic model in order to explore, e.g., system-level implications of local perturbations. The goal is to keep the basic core model compact enough to insure fast execution time (in view of eventual use in the clinic) and yet to allow elaborate detailed modules of target tissues or organs in order to focus on the problem area while maintaining the system-level regulatory compensations.

Development of the modeling environment is proceeding in several stages in order to provide practical tools even in the early phases of the project. The basic system variables of the core model, chosen to reflect the needs of the initial target problems, are blood pressure (both short- and long-term regulation), pH, concentrations of principal solutes, and the volumes of the various fluid compartments. The essential regulatory variables are autonomic tone and the levels of the hormones vasopressin, aldosterone, angiotensin. The core model includes basic descriptions of the heart, vasculature, intra- and extracellular spaces, lungs, kidneys, and muscles. Rather than starting from scratch, the core model is based on two legacy models that treated overall regulation of blood pressure (Guyton et al. 1972) and fluid regulation (Ikeda et al. 1979).

The resulting modeling resource will be provided as an Open Source project and will be made available to the general community as part of the IUPS Physiome effort.

In parallel to the model development, the project includes a datawarehouse for the vast collection of heterogeneous experimental data necessary not only for evaluation of the many parameter values but also for experimental and clinical validation of the simulation results. This will involve ontology development and implementation of grid computing systems-mediation software.

Manuscript received April 16, 2007. This work was supported in part by the French National Research Agency (SAPHIR, ANR-06-BYOS-0007-01) and by the project FAME2, http://www.fame2.org/, in the competitivity cluster SYSTEM@TIC PARIS-REGION, http://www.systematic-paris-region.org/.

S. R. Thomas, P. Maziere, F. Tahi, and F. Zehraoui are with the CNRS, FRE 2873 IBISC (Informatics, Integrative Biology and Complex Systems), in Evry, France (phone: +33-1-60873928; fax: +33-1-60873789; e-mail: srthomas@ibisc.fr).
P. Hannaert and F. Guillaud are with Inserm E0324 (Ischémie-reperfusion en transplantation rénale), Poitiers.
P. Baconnier, E. Abdulhay and J. Fontecave are with UMR CNRS 5525 TIMC-IMAG, Grenoble.
A. Hernandez and V. Le Rolle are with LTSI, INSERM U-642, Rennes.
J-P Françoise is with Univ. Paris VI, Paris.

## II. METHODS AND PRELIMINARY RESULTS

### A. Core model development

The core systems model includes basic input-output, reduced-dimension formal representations of the essential components involved in whole body regulation of blood pressure and body fluids, namely: renal function, respiration, intra- and extracellular fluid spaces, and cardiovascular subsystems. Also included are the main regulatory sensors (baro- and chemoreceptors) and nervous (autonomic control) and hormonal regulators (vasopressin, aldosterone, angiotensin). While models have previously been developed with a similar systems approach, ours is the first that aims to provide an open-source, interactive, modular resource for the general research community, amenable to plug-and-play exploration of the systems implications of detailed and focused models of individual subsystems. All aspects of the model will be open to inspection and modification by users of the resource. We will produce scaled versions of the basic core model for human, of course, and also for rat, mouse, and dog, since these are the principle experimental models.

Two previous integrated models have served as the starting point for development of the core model, namely, the classic model of Guyton, Coleman, & Granger [1], which focused on blood pressure regulation, and the model of Ikeda et al. [2] which focused on overall regulation of body fluids. We have functioning re-implementations of the Guyton et al. and Ikeda et al. models in Fortran, C++, and also using the ODE solver Berkeley Madonna. The actual parameter values and functional relationships in our core model are of course adjustable to take into account the 30 years of progress since publication of the original models. The set of parameter values actually used will be clearly documented, and all related work from the literature will be collected in a Quantitative Systems DataBase (QSDB) which we will build from our generic QxDB/QKDB database environment [3, 4].

The basic system variables of the core model, chosen to reflect the needs of the initial target problems, are blood pressure (both short- and long-term regulation), concentrations of (at a minimum) NaCl, glucose, lactate, urea, $PO_2$, $PCO_2$, $HCO_3^-$, $NH_3/NH_4^+$, and pH, and the volumes of the various compartments. The essential regulatory variables are autonomic tone and the levels of the hormones vasopressin, aldosterone, angiotensin, and insulin.

The core model will be accessible via an interactive web interface and also in several other formats, such as Matlab, Mathematica, and Berkeley Madonna (this list may be extended later).

*B. Modeling and simulation tools, database & ontologies & grid infrastructure*

As stated above, the core model consists of a number of modules, each having clearly specified input and output variables (or arguments). Within each module, an appropriate submodel relates the input variables to the output variables, either by some simple transfer function or by a more-or-less detailed model of the subsystem involved. In the latter case, the sub-module models may be developed in any of a number of different formalisms, e.g., differential-algebraic equations, multi-agent systems, cellular automata, etc. Consistent with the object-oriented spirit of the core modeling environment, the only restriction on each sub-module is that it respect the imposed input/output characteristics in order to correctly interface with the rest of the systems model. Note that within a given submodule, "local" variables may be defined, permitting to follow, during actual *in silico* simulations, any additional appropriate processes taking place within the considered subsystem. For example, a hybrid (multi-formalism) electrophysiological model of an ischemic cardiac tissue, can be constructed as a set of coupled sub-models representing healthy and ischemic cells [5, 6]. Healthy cells can be represented by means of a simplified (discrete) automaton model, while ischemic cells can be represented with a detailed (continuous) model. We can choose to record only the state variables of some ischemic cells (local variables of the continuous submodels) and the global activity at the tissue level (the EGM potential generated by the set of cells), as the output of the global model.

Technically, this approach requires a robust and flexible simulator architecture, in order to keep the many processes in step with each other while minimizing computation time. Numerical techniques must also be appropriate for the "stiff" nature of the problem, since some processes are orders of magnitude faster than others in this system (e.g., in the above example, a fourth-order RK integration method with adaptive step-size is used for detailed models, while a discrete-time simulator is used for simplified models). The LTSI/SEPIA team (Rennes) has developed such an environment, called M2SL, based on the co-simulation principle [7] and has demonstrated its use for several problems in cardiovascular physiopathology and epileptology [8-10].

These techniques scale up nicely to the systems model level of the present project.

*Database & ontology development, & grid architecture.* We are implementing a datawarehouse and a set of ontologies in order to accomodate integration of renal, cardiac and respiratory data. The data we are mostly concerned with are physiological data needed for parameter estimation and clinical data needed for model validation and benchmarking. As development of detailed sub-modules proceeds, this will extend to cellular and molecular level information such as kinetic descriptions of membrane channels and cotransport proteins.

This infrastructure is being built as an extension of QKDB (http://physiome.ibisc.fr/qkdb/), the Quantitative Kidney Database.

*C. Development of detailed sub-modules*

In parallel with development of the core model, we are building a collection of detailed submodules for the major organ systems involved in blood pressure regulation and systems-level fluid and solute homeostasis. This involves adaptation of legacy, published models of the heart, lung, and kidney at many scales and at various levels of resolution. The adaptation consists of implementation in the 'language' of the SAPHIR toolbox, with standardized specification of input and output variables.

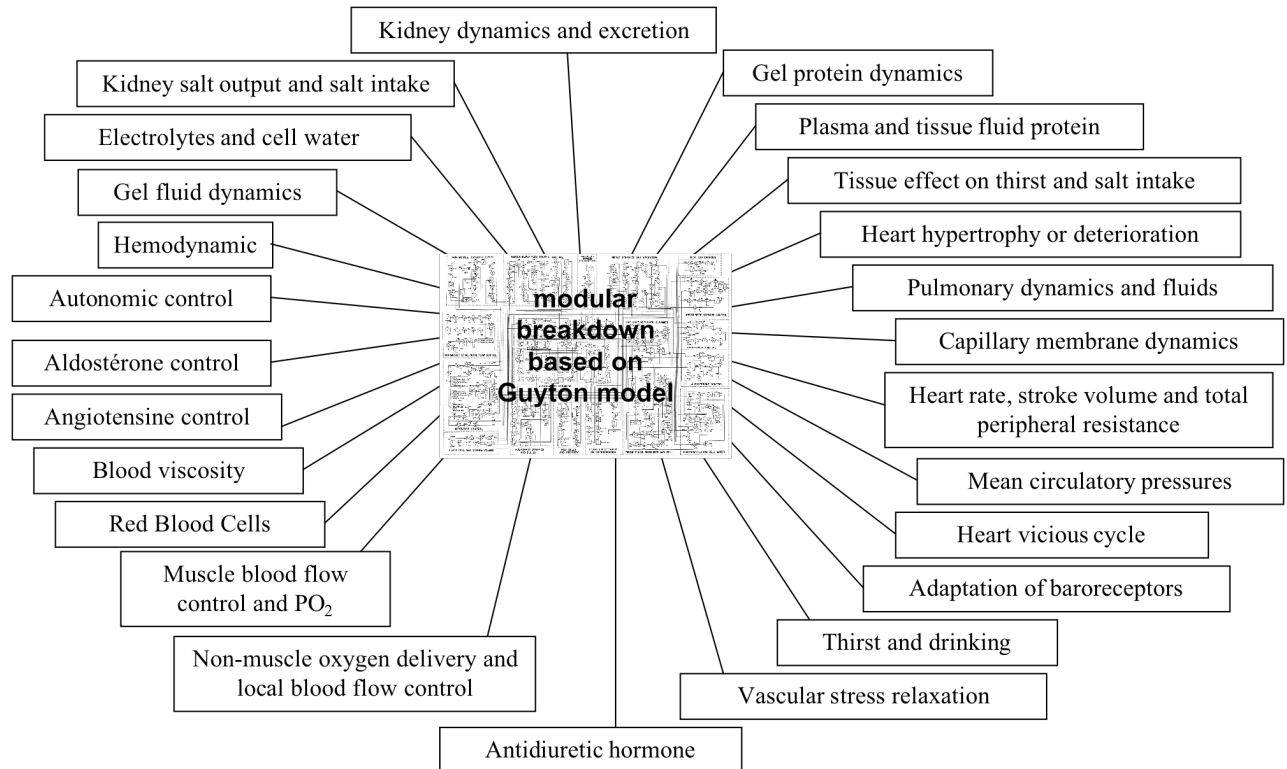

Figure 1. Diagram of the modular multi-scale modeling environment. Using the Guyton model as starting point, we defined a number of object-oriented sub-modules. The whole collection is solved using the M2SL toolbox (Rennes lab), which automatically finds optimum step-size for individual modules and distributes the calculation over a number of parallel processors, when available.

*Renal module:* The kidney is the main actor in long-term regulation of blood pressure and hydro-electrolytic equilibrium [11-14]. As such, it plays, for example, a pathogenic role in most forms of hypertension: hypertension accelerates diseased kidney function loss, whereas antihypertensive treatment slows renal diseases progression.

The primary role of the kidney module (KM) is to bestow the whole-body model with kidney-dependent functionality related to regulation of blood pressure, volemia, and sodium balance [15,16]. At a minimum, the KM will implement glomerular filtration, tubule $Na^+$ & water transport, tubulo-glomerular feedback, renin production, hormone actions, urine concentration in the renal medulla, and consequent natriuresis and diuresis. Additional factors might include endothelin, catecholamines, and dopamine.

*Cardiovascular module:* The analysis of autonomic nervous system (ANS) activity and, particularly, of the way it modulates the cardiovascular system (e.g. analysis of heart rate variability or ventricular contractility) has been shown to provide valuable information for risk-stratification and early detection of cardiovascular pathologies such as cardiac ischemia and heart failure [17, 18]. In clinical practice, the analysis of the ANS is commonly performed by applying a set of tests called autonomic manoeuvres (such as the Valsalva manoeuvre [19] or the Tilt test [20]). These manoeuvres are based on the controlled modification of a single cardiovascular variable in order to observe the regulatory response of the ANS. Various observations, such as the electrocardiogram (ECG), the noninvasive systemic arterial pressure (SAP) or respiration, are acquired concurrently, to better characterize this autonomic response. However, the interpretation of these data can be very difficult, due to the multidimensionality of the observed phenomena and the fact that the complex mechanisms involved in the autonomic regulation of the cardiovascular system are not fully understood. For example, the reduced blood flow and arterial pressure observed during heart failure lead to chronic sympathetic activation, decreased parasympathetic activity and impaired arterial baroreflex activity. These abnormalities in the autonomic control of the cardiovascular system are associated with pronounced heart failure, higher mortality rates, and cardiac arrhythmia. A model-based analysis of this pathology seems particularly appropriate, since it will help to take into account the variety of system-level functions.

To improve the core model, detailed submodules will eventually include more detailed models of: i) cardiac mechanical activity, ii) the arterial and venous vascular systems, iii) autonomic baroreflex loop including afferent and efferent pathways (short-term regulatory loop), and iv) the autonomic long-term regulatory loop.

A multi-resolution model of the electro-mechanical activity of the heart will enable simulation of the consequences of a local ventricular desynchronization on the function of the whole system.

Also underway is identification of model parameters from clinical observations: A database of cardio-respiratory signals is in construction in collaboration with the University Hospital of Rennes, by applying Tilt tests and Valsalva maneuvers to a population of normal subjects and patients suffering from heart failure. Observations include: 12-leads ECG, continuous non-invasive blood pressure, impedance cardiography and a complete echocardiographic analysis. These data will also be recorded in GDF format and included in the database application mentioned above.

Previous models of cardiac electrical and mechanical activities [8, 9, 21, 23] will be used as reference to create the new coupled multi-resolution model.

*Lung module:* Over the time scale of the respiratory rhythm, the effect of blood pressure changes on respiratory mechanics is rather well described. However, this is not the case at the cardiac frequency scale. Indeed, the effect of pleural pressure on the heart is quite well elucidated [22]. Nonetheless, to our knowledge, the reciprocal effect of heart on pleural pressure is not clear: the beating heart can be considered as a volume and/or pressure change generator inside the rib cage as evidenced by its effect on lung volume [23]. The Grenoble team is investigating this closed-loop mechanical interaction. Sensitivity analysis and parameter identification will be carried out using data obtained on healthy volunteers. Recorded physiological variables do not imply recourse to invasive techniques: airway flow, thoracic volume changes, and mouth pressure at occlusion give enough information for the parameter identiification process.

*Central cardio-respiratory interactions:* Cardio-respiratory interactions are well documented on an experimental and pathological basis. A simple model of cardio-respiratory coupling [22, 24] is being translated into the project formalism. At this stage, the simpler the better, since the first question is: what are the possible and effective mechanical and central cardio-respiratory interactions? An integrated model including both central and mechanical interactions should provide an answer to this typical integrative physiology question.

Users of the environment will be able to assemble a customized collection of core-modules and detailed sub-modules, depending on the question they wish to address.

I. CONCLUSION

This modular, comprehensive systems modeling resource will be an important step toward the general goal of the IUPS Physiome project to eventually attain an overall "gene-to-organism" modeling environment. One of the principal, immediate goals will be to contribute to improved management of hypertension, through more accurate prediction of antihypertensives such as certain diuretics, and to better understand the complex interplay of short- and long-term regulations.

ACKNOWLEDGMENT

The authors thank R. Garay for his help and expertise in hypertension. We also are greatly indebted to Ron White for his inestimable help in the re-implementation of the original models of Guyton and colleagues.